\documentclass[twoside,twocolumn,9pt]{article}
\usepackage{extsizes}
\usepackage[super,sort&compress,comma]{natbib} 
\usepackage[version=3]{mhchem}
\usepackage[left=1.5cm, right=1.5cm, top=1.785cm, bottom=2.0cm]{geometry}
\usepackage{balance}
\usepackage{times,mathptmx}
\usepackage{sectsty}
\usepackage{graphicx} 
\usepackage{lastpage}
\usepackage[format=plain,justification=justified,singlelinecheck=false,font={stretch=1.125,small,sf},labelfont=bf,labelsep=space]{caption}
\usepackage{float}
\usepackage{fancyhdr}
\usepackage{fnpos}
\usepackage[english]{babel}
\addto{\captionsenglish}{%
  
}
\usepackage{array}
\usepackage{droidsans}
\usepackage{charter}
\usepackage[T1]{fontenc}
\usepackage[usenames,dvipsnames]{xcolor}
\usepackage{setspace}
\usepackage[compact]{titlesec}
\usepackage{graphicx}% Include figure files
\usepackage{dcolumn}% Align table columns on decimal point
\usepackage{bm}% bold math
\usepackage{txfonts}
\usepackage{braket}
\usepackage{mathtools}
\usepackage{epstopdf}%This line makes .eps figures into .pdf - please comment out if not required.

\definecolor{cream}{RGB}{222,217,201}

\begin{document}

\pagestyle{fancy}
\thispagestyle{plain}
\fancypagestyle{plain}{

%%%HEADER%%%
\fancyhead[C]{\includegraphics[width=18.5cm]{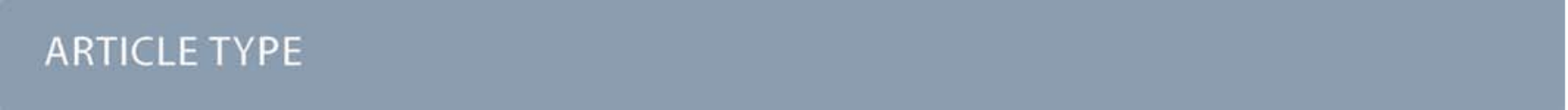}}
\fancyhead[L]{\hspace{0cm}\vspace{1.5cm}\includegraphics[height=30pt]{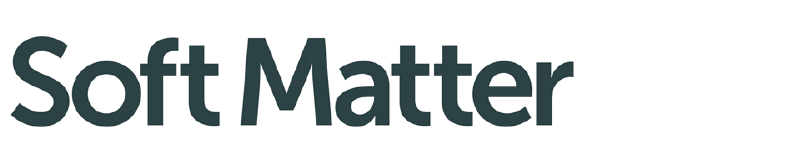}}
\fancyhead[R]{\hspace{0cm}\vspace{1.7cm}\includegraphics[height=55pt]{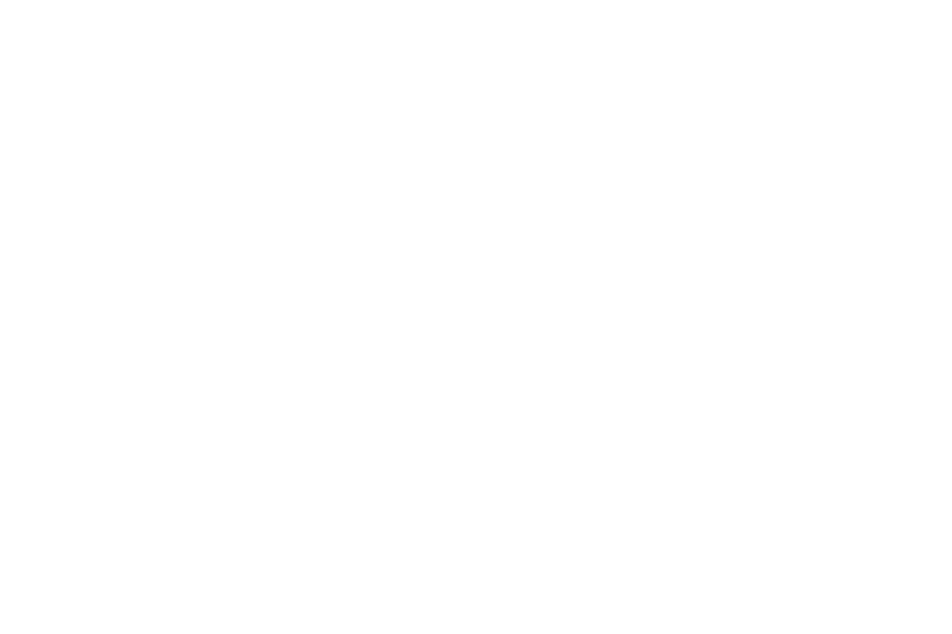}}
\renewcommand{\headrulewidth}{0pt}
}
%%%END OF HEADER%%%

%%%PAGE SETUP - Please do not change any commands within this section%%%
\makeFNbottom
\makeatletter
\renewcommand\LARGE{\@setfontsize\LARGE{15pt}{17}}
\renewcommand\Large{\@setfontsize\Large{12pt}{14}}
\renewcommand\large{\@setfontsize\large{10pt}{12}}
\renewcommand\footnotesize{\@setfontsize\footnotesize{7pt}{10}}
\makeatother

\renewcommand{\thefootnote}{\fnsymbol{footnote}}
\renewcommand\footnoterule{\vspace*{1pt}% 
\color{cream}\hrule width 3.5in height 0.4pt \color{black}\vspace*{5pt}} 
\setcounter{secnumdepth}{5}

\makeatletter 
\renewcommand\@biblabel[1]{#1}            
\renewcommand\@makefntext[1]% 
{\noindent\makebox[0pt][r]{\@thefnmark\,}#1}
\makeatother 
\renewcommand{\figurename}{\small{Fig.}~}
\sectionfont{\sffamily\Large}
\subsectionfont{\normalsize}
\subsubsectionfont{\bf}
\setstretch{1.125} %In particular, please do not alter this line.
\setlength{\skip\footins}{0.8cm}
\setlength{\footnotesep}{0.25cm}
\setlength{\jot}{10pt}
\titlespacing*{\section}{0pt}{4pt}{4pt}
\titlespacing*{\subsection}{0pt}{15pt}{1pt}
%%%END OF PAGE SETUP%%%

%%%FOOTER%%%
\fancyfoot{}
\fancyfoot[LO,RE]{\vspace{-7.1pt}\includegraphics[height=9pt]{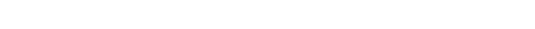}}
\fancyfoot[CO]{\vspace{-7.1pt}\hspace{13.2cm}\includegraphics{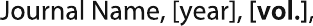}}
\fancyfoot[CE]{\vspace{-7.2pt}\hspace{-14.2cm}\includegraphics{head_foot/RF}}
\fancyfoot[RO]{\footnotesize{\sffamily{1--\pageref{LastPage} ~\textbar  \hspace{2pt}\thepage}}}
\fancyfoot[LE]{\footnotesize{\sffamily{\thepage~\textbar\hspace{3.45cm} 1--\pageref{LastPage}}}}
\fancyhead{}
\renewcommand{\headrulewidth}{0pt} 
\renewcommand{\footrulewidth}{0pt}
\setlength{\arrayrulewidth}{1pt}
\setlength{\columnsep}{6.5mm}
\setlength\bibsep{1pt}
%%%END OF FOOTER%%%

%%%FIGURE SETUP - please do not change any commands within this section%%%
\makeatletter 
\newlength{\figrulesep} 
\setlength{\figrulesep}{0.5\textfloatsep} 

\newcommand{\topfigrule}{\vspace*{-1pt}% 
\noindent{\color{cream}\rule[-\figrulesep]{\columnwidth}{1.5pt}} }

\newcommand{\botfigrule}{\vspace*{-2pt}% 
\noindent{\color{cream}\rule[\figrulesep]{\columnwidth}{1.5pt}} }

\newcommand{\dblfigrule}{\vspace*{-1pt}% 
\noindent{\color{cream}\rule[-\figrulesep]{\textwidth}{1.5pt}} }

\makeatother
%%%END OF FIGURE SETUP%%%

%%%TITLE, AUTHORS AND ABSTRACT%%%
\twocolumn[
  \begin{@twocolumnfalse}
\vspace{3cm}
\sffamily
\begin{tabular}{m{4.5cm} p{13.5cm} }

\includegraphics{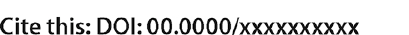} & \noindent\LARGE{\textbf{Mixing--Demixing Transition in Polymer-Grafted Spherical Nanoparticles}} \\
\vspace{0.3cm} & \vspace{0.3cm} \\

% & \noindent\large{Peter Yatshyshin,$^{\ast}$\textit{$^{a}$} Nikolaos G. Fytas,\textit{$^{b\ddag}$} and Panagiotis E. Theodorakis\textit{$^{a}$}} \\%Author names go here instead of "Full name", etc.
 & \noindent\large{Peter Yatsyshin,$^{\ast}$\textit{$^{a}$} Nikolaos G. Fytas,$^{\ast}$\textit{$^{b}$} and Panagiotis E. Theodorakis$^{\ast}$\textit{$^{c}$}} \\%Author names go here instead of "Full name", etc.

\includegraphics{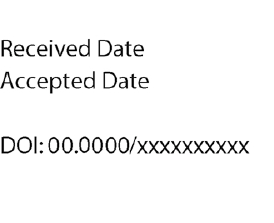} & \noindent\normalsize{Polymer-grafted nanoparticles (PGNPs) can provide property profiles than cannot be obtained
individually by polymers or nanoparticles (NPs). Here, we have studied the mixing--demixing
transition of symmetric copolymer melts of polymer-grafted spherical nanoparticles
by means of coarse-grained molecular dynamics simulation and a theoretical
mean-field model. 
We find that a larger size of NPs leads to higher stability for given 
number of grafted chains and chain length reaching a point where demixing is 
not possible. Most importantly, the increase
in the number of grafted chains, $N_g$, can initially 
favour the phase separation of PGNPs, but
further increase can lead to more difficult demixing. The reason is
the increasing impact of an effective  core that forms as
the grafting density of the tethered polymer 
chains around the NPs increases. The range and exact values of $N_g$ where 
this change in behaviour takes place depends on the NP size
and the chain length of the grafted polymer chains. 
Our study elucidates the phase behaviour of PGNPs and in particular the influence of the grafting density on the phase behaviour of the systems
anticipating that it will open new doors in the
understanding of these systems with 
implications in materials science and medicine.} \\%The abstrast goes here instead of the text "The abstract should be..."

\end{tabular}

 \end{@twocolumnfalse} \vspace{0.6cm}

  ]
%%%END OF TITLE, AUTHORS AND ABSTRACT%%%

%%%FONT SETUP - please do not change any commands within this section
\renewcommand*\rmdefault{bch}\normalfont\upshape
\rmfamily
\section*{}
\vspace{-1cm}

%%%FOOTNOTES%%%

\footnotetext{\textit{$^{a}$~Department of Chemical Engineering, Imperial College London, South Kensington Campus, SW7 2AZ London, UK; E-mail: p.yatsyshin@imperial.ac.uk}}
\footnotetext{\textit{$^{b}$~Applied Mathematics Research Centre, Coventry University,\ 
Coventry CV1 5FB, United Kingdom; E-mail: nikolaos.fytas@coventry.ac.uk }}
\footnotetext{\textit{$^{c}$~Institute of Physics, Polish Academy of Sciences, Al.\
Lotnik\'ow 32/46, 02-668 Warsaw, Poland; E-mail: panos@ifpan.edu.pl }}

%Please use \dag to cite the ESI in the main text of the article.
%If you article does not have ESI please remove the the \dag symbol from the title and the footnotetext below.
%\footnotetext{\dag~Electronic Supplementary Information (ESI) available: [details of any supplementary information available should be included here]. See DOI: 10.1039/cXsm00000x/}
%additional addresses can be cited as above using the lower-case letters, c, d, e... If all authors are from the same address, no letter is required

%\footnotetext{\ddag~Additional footnotes to the title and authors can be included \textit{e.g.}\ `Present address:' or `These authors contributed equally to this work' as above using the symbols: \ddag, \textsection, and \P. Please place the appropriate symbol next to the author's name and include a \texttt{\textbackslash footnotetext} entry in the the correct place in the list.}

%%%END OF FOOTNOTES%%%

%%%MAIN TEXT%%%%
\section{Introduction}
Inorganic nanoparticles (NPs) dispersed in 
polymer hosts have attracted much attention 
over the last decades in multiple technological areas, such as electronics, medicine,
and others \cite{Balazs2006,Choudhury2015,Schaefer2011}. In these applications,
nanocomposite materials have
property profiles that cannot be obtained by using polymers or NPs alone, 
for example, NPs in a polymer host must avoid aggregation. However,
creating homogeneous mixtures of NPs and polymers turns out to be 
challenging, due to the attractive van der Waals forces between NPs and
the polymer-mediated depletion interactions \cite{Hooper2005,Hooper2006}. 
A possible solution to this problem is 
NPs grafted with polymer chains \cite{Sunday2012,Srivastava2017,Srivastava2014,Mangal2015,Srivastava2012}, 
known as polymer-grafted or polymer-tethered NPs (PGNPs), which can be self-suspended with
homogeneous particle dispersion in the absence of any
solvent \cite{Agrawal2016,Chremos2011,Chremos2011b,Chremos2016,Chremos2017}. 
Density functional theory
has indicated that the phase stability is due to the space-filling constraint on the grafted
corona chains \cite{Yu2010,Yu2014,Chremos2011b}, which leads to an 
effective attraction of entropic origin between NPs, which
is also mediated by the attached polymer chains. In this case, the Flory--Huggins parameter 
may be larger and more negative \cite{Yu2014,Agarwal2010}. 
Another reason for the steric stabilization of the NPs
may be the absence of the solvent and the 
presence of incompressible polymer chains. This results
in the suppression of long wavelength fluctuations
that induces an effective attraction 
between the particles \cite{Yu2010,Yu2014,Agarwal2010,Fernandes2013,Chremos2011b}. 
The constraints on the
corona imposed by space-filling have been in the centre of attention of recent work, and in
particular, the dependence on the core particle size
\cite{Agrawal2015,Chremos2011,Chremos2016,Chremos2017,Chremos2011b}, chain length
\cite{Kim2015,Agarwal2012,Chremos2011b,Chremos2016}, grafting density
\cite{Kim2015,Kim2014,Choudhury2015,Choudhury2015b,Goyal2011,Chremos2011,Chremos2011b}, and
temperature \cite{Choudhury2015,Agarwal2011,Chremos2011}. Recently,
there has been also studies of
PGNPs in blends with chemically distinct polymers focusing on the physical properties,
structure, and underlying dynamics of these systems \cite{Agrawal2016}. 
This latter work has been corroborated 
with theoretical results that allowed the estimation of the heat of mixing of the PGNP blends
as a function of the volume fraction of the system \cite{Agrawal2016}. 

A theoretical description of the phase behaviour for PGNPs
mixtures based on arguments of the Flory--Huggins 
theory \cite{Meyer1940,Flory:1953/a} is challenging, even for the most symmetric cases in
composition and molecular architecture. This is due to the fact that the theoretical 
assumptions related to a necessary
effective $\chi_{\rm eff}$ parameter that describes systems 
of PGNPs
require extensive testing and validation with experimental 
and simulation results. This effective parameter would correspond
to an effective chain length $N_{\rm eff}$ of effective polymer chains, 
$\chi_{\rm eff} \sim N_{\rm eff}$. Moreover,  $\chi = \alpha/T + \beta$,
where $\alpha$ and $\beta$ are parameters
for the enthalpic and entropic contributions to $\chi$ \cite{Chremos:2014/a}. These
parameters are controlled by the interaction between 
the different components. 
In addition, the interplay between entropic and enthalpic contributions is affected in a nontrivial manner by parameters 
pertaining to the architecture of PGNPs. 
As a result, the theoretical prediction of $\chi$ for these systems is far from being under command.

In this work, we address this issue by studying the order--disorder transition (ODT) of
composition- and molecular-symmetric spherical PGNP melts, \textit{i.e.}, the attached polymer
chains are all of the same length, NPs are spherical and of the same size and the same
grafting density, but grafted polymer chains differ in their chemical type.
The symmetry of our system will allow us to isolate main features that
dictate the phase behaviour of PGNPs.
By using molecular dynamics (MD) simulations of a bead--spring 
model \cite{Kremer1990,Chremos2011,Chremos2014panos,ChremosTheodorakis2016}, we 
demonstrate that the variation of the NPs size 
leads to higher stability of the melts for given grafting density 
and length of the polymer chains. For small and intermediate 
size of NPs we find a transition region that reflects
a change in the phase behaviour of the system. Initially, the increase in the number of grafted chains favours the separation of the PGNPs. However, further increase renders the demixing of PGNPs more difficult for a given polymer length and NP size. One possible explanation of this behaviour is the increase of the effective core size of PGNPs that is due to the increase in the number of the tethered chains around the spherical core particle. The range and location of this change in behaviour depends on the NP size and chain length. A simple understanding of these findings can be gained in terms of an intuitive mean-field model, where the PGNP melt is coarse-grained as a fluid mixture, with components having repulsive interactions. 
Overall, our results describe the mixing/demixing tendencies of 
symmetric PGNPs melts elucidating the specific effects of PGNPs core, which are at the heart of the PGNPs molecular architecture for moderate size of NPs and grafted polymer chains. In this way, we anticipate that our findings will provide a fundamental understanding towards the 
design of PGNPs structures, which is important for advanced applications 
in materials science, medicine, and beyond.  

\begin{figure}
\vspace{2mm}
\includegraphics[width=90mm]{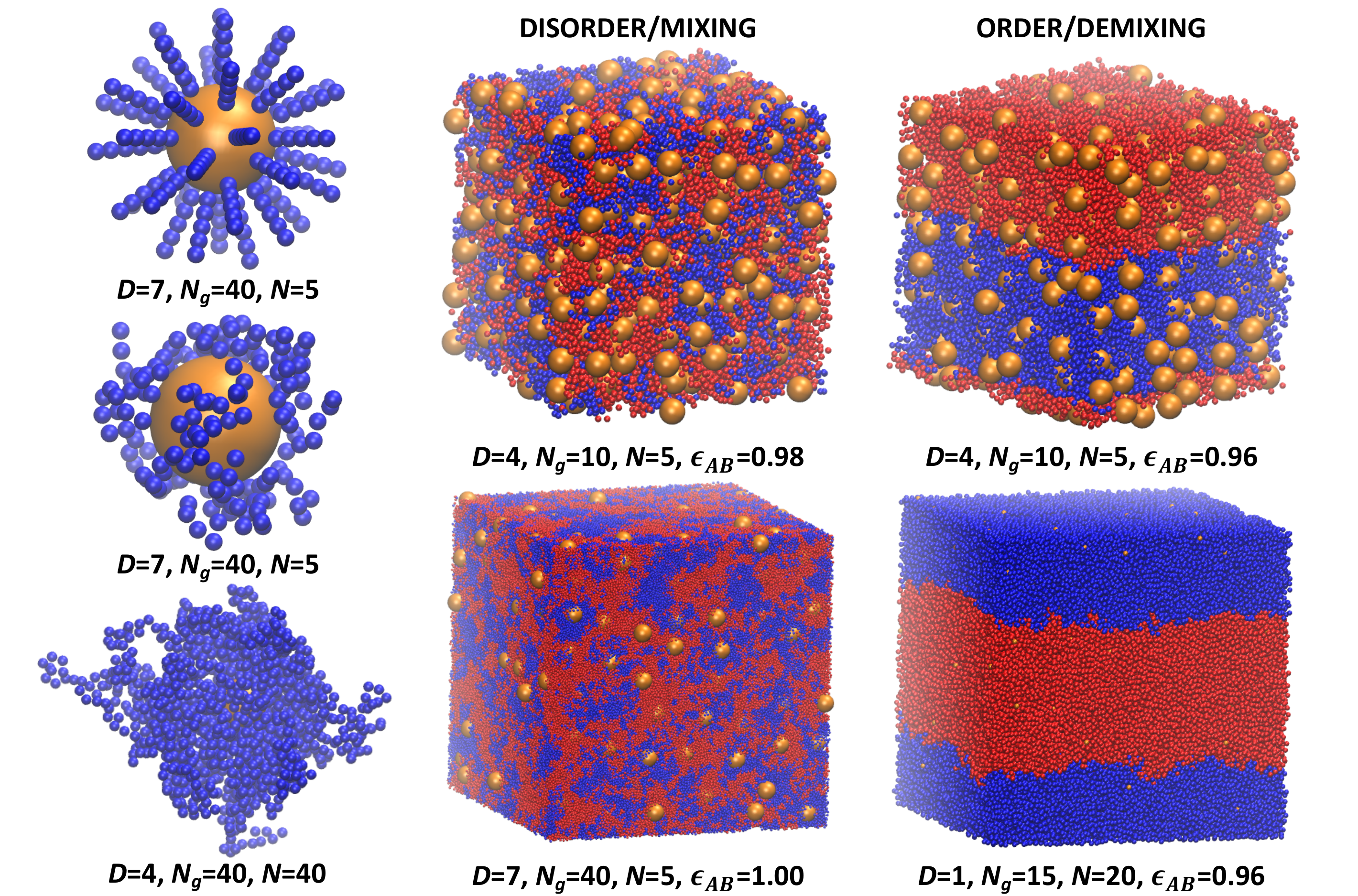}
\caption{Left panel from top to bottom depicts the molecular architecture of a PGNP, 
which is used to construct the initial configuration
of the system, and two cases 
(NP is covered or not by the grafted polymer chains)
as they appear in different melts for different parameters, as
indicated. The rest of the figure illustrates examples 
of mixed (disorder) and demixed (order) melts for 
different sets of parameters as indicated. $D$ is the size of the
core NP, $N_{\rm g}$ is the number of grafted chains, 
$N$ is the length of each tethered polymer chain, 
and $\epsilon_{AB}$ expresses the incompatibility 
between red and blue beads (smaller values of $\epsilon_{AB}$ 
correspond to larger incompatibility). The temperature for all systems
was $T=0.8 \epsilon/k_B$.}
\label{Fig1}
\end{figure}

\section{Methodology}
%\vspace{0.5cm}
%\noindent
We implemented MD simulations of a standard bead-spring 
model in the NPT ensemble by using the large-scale atomic/molecular massively parallel 
simulator (LAMMPS) \cite{Plimpton:1995/a}. The pressure and
temperature of the system was fluctuating around predefined values during
the course of the simulation. In our case, the temperature was 
$T=0.8\epsilon/k_B$ with 
$\epsilon$ being the unit of energy and $k_B$ the Boltzmann
constant, while the pressure ($P=0.1 \epsilon/\sigma^3$, where $\sigma$ is the unit of length) corresponded to the 
ambient pressure and is the value used in previous studies \cite{Chremos2011,Chremos2017,Chremos2016,Chremos2011b}.  Each simulated PGNPs copolymer melt consists of 500
molecules placed in a cubic simulation box with periodic boundary conditions applied in
all Cartesian directions. Each molecule is composed of a spherical core
particle with diameter, $D$, which are labelled in our model with the index `c'. 
Here, we have considered the cases $D=1\sigma,4\sigma$,
and $7\sigma$, which are suitable for our 
simulations \cite{Chremos2011,Chremos2011b,Chremos2016,Chremos2017}. Linear
polymer chains of either type A or B monomers are grafted to each core NP with 
grafting density $d_{g}=N_{g}/A_c$, where $A_{c}=\pi D^2$ is the surface area of the 
core particle in units of $\sigma^2$, and $N_{g}$ is the number of grafted chains. The number of
grafted chains, which is the same for all PGNPs in the melt,  in this study ranges
from $5$ to $40$ when $D=4\sigma$ or $7\sigma$, while for $D=1\sigma$ the maximum number
of grafted chains is $N_{g}=15$, due to the small size
of the core NP. A previous study\cite{Chremos2011} has determined that the possible
maximum number of grafted chains for the $D=1\sigma$ case is $N_{g}=16$. 
Moreover, grafted polymer chains are homogeneously
distributed on the NP's surface by using the Fibonacci lattice (Fig.~\ref{Fig1}, 
upper left panel). 
The length of the polymer chains
is $N$, which ranges from $5$ to $40$. Thus, our melts are symmetric in 
architecture and composition, namely each NP has the same number of polymer
chains with the same length, while half of these PGNPs have type A grafted polymer chains
and the other half has NPs with tethered polymers of type B. Beads of type c, A, and B interact 
by means of the Lennard--Jones (LJ) potential
\begin{equation}\label{eq:LJpotential}
U_{\rm LJ}=4\epsilon_{\rm ij} \left[  \left(\frac{\sigma_{\rm ij}}{r}  \right)^{12} -    \left(\frac{\sigma_{\rm ij}}{r}  \right)^{6}    \right],
\end{equation}
where $r_{\rm ij}$ is the distance between any beads of type $\rm i$ 
and $\rm j$. The potential is cut
and shifted with the cutoff distance for the interactions between polymer beads
being $r_{\rm c}=2.5\sigma_{\rm ij}$. The cutoff distance between any pair of interactions 
that involve the core particle
is set to $2^{1/6}\sigma_{\rm ij}$. The potential parameters in our case are
$\sigma_{\rm AA}=\sigma_{\rm BB}=\sigma_{\rm AB}=\sigma$ and $\epsilon_{\rm AA}=\epsilon_{\rm BB}=\epsilon$.
In our study, the size of the spherical NPs varied, namely we explored the
cases $D=\sigma_{\rm cc}=\sigma, 4\sigma$ and $7\sigma$. Also, the 
cross-interaction $\epsilon_{\rm AB}$ between A and B beads, which is used to cross the 
order--disorder transition, typically varied between 
$0.5\epsilon$ and $1.0\epsilon$. Hence, the degree of incompatibility between
type A and B polymer chains is introduced 
by varying $\epsilon_{\rm AB}$ \cite{Chremos:2014/a}. 
Polymer chains are fully flexible and they are connected by
harmonic bonds, namely 
$V_{\rm h}=k(r-r_{0})^2$, where $r_{0}=\sigma$ is the equilibrium bond length and 
$k=10000\epsilon/\sigma^2$ is a constant. Grafted beads are
immobile on the surface of the core NP. Depending on the particular 
parameters, we run our simulations up to $10^9$ MD times steps, with each time
step corresponding to $\delta t= 0.005\tau$, where 
$\tau=\sigma(m/\epsilon)^{1/2}$ is the natural time unit with $m$ indicating the unit of mass. Each simulation
at a lower $\epsilon_{\rm AB}$ is a continuation of a previous simulation
at higher $\epsilon_{\rm AB}$. This procedure was employed for different
initial snapshots. Moreover, we did not observe any
hysteresis effects when we repeatedly increased or decreased the value
of $\epsilon_{AB}$ close to the demixing point. Given previous studies
of these systems that suggest a star-forming liquid behaviour for
$D>5\sigma$ and $N_g>60$, we can assume that our conclusions are not 
affected by kinetic limitations in our systems \cite{Chremos2011}.

\begin{figure}[t]
\includegraphics[width=90mm]{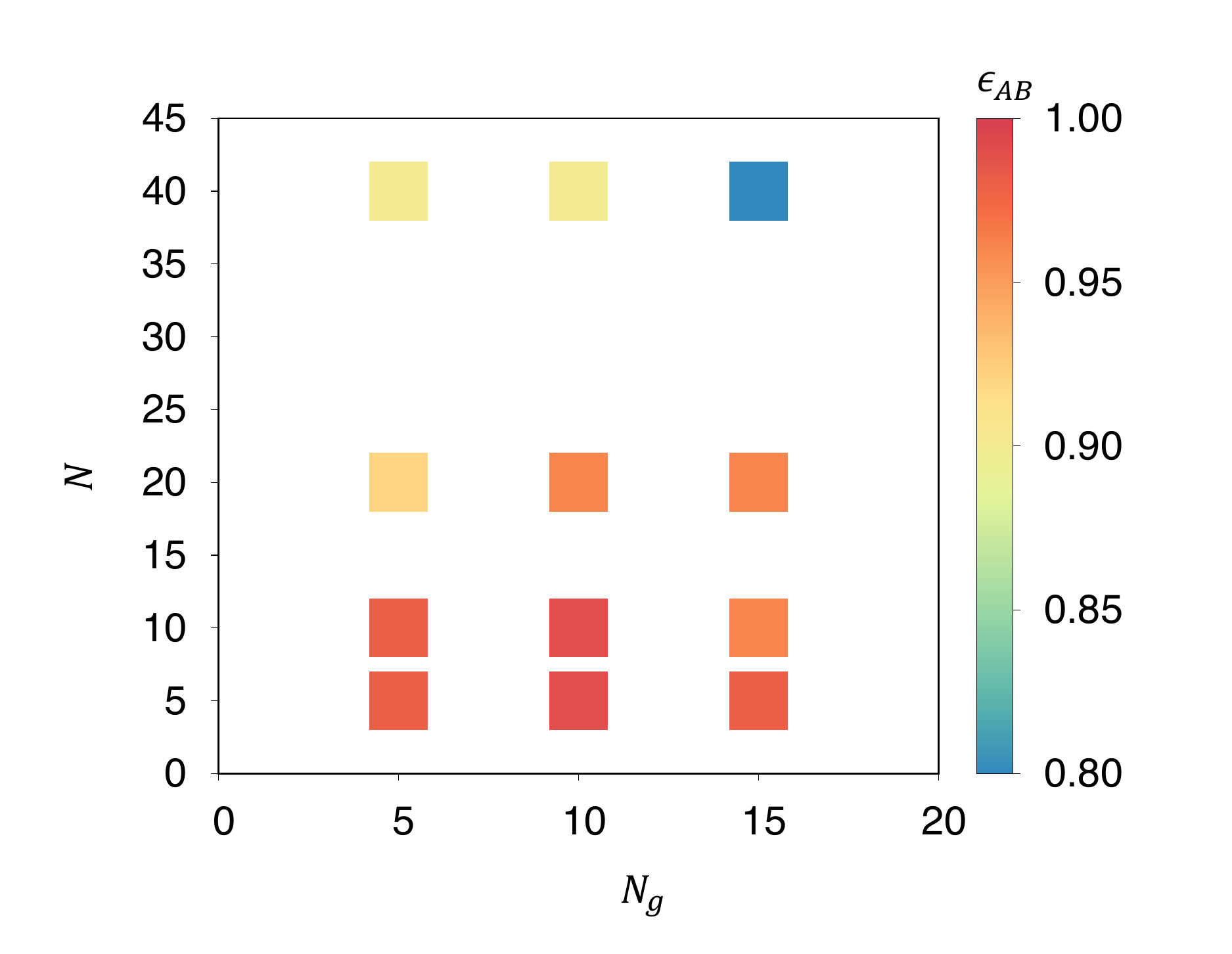} 
\includegraphics[width=90mm]{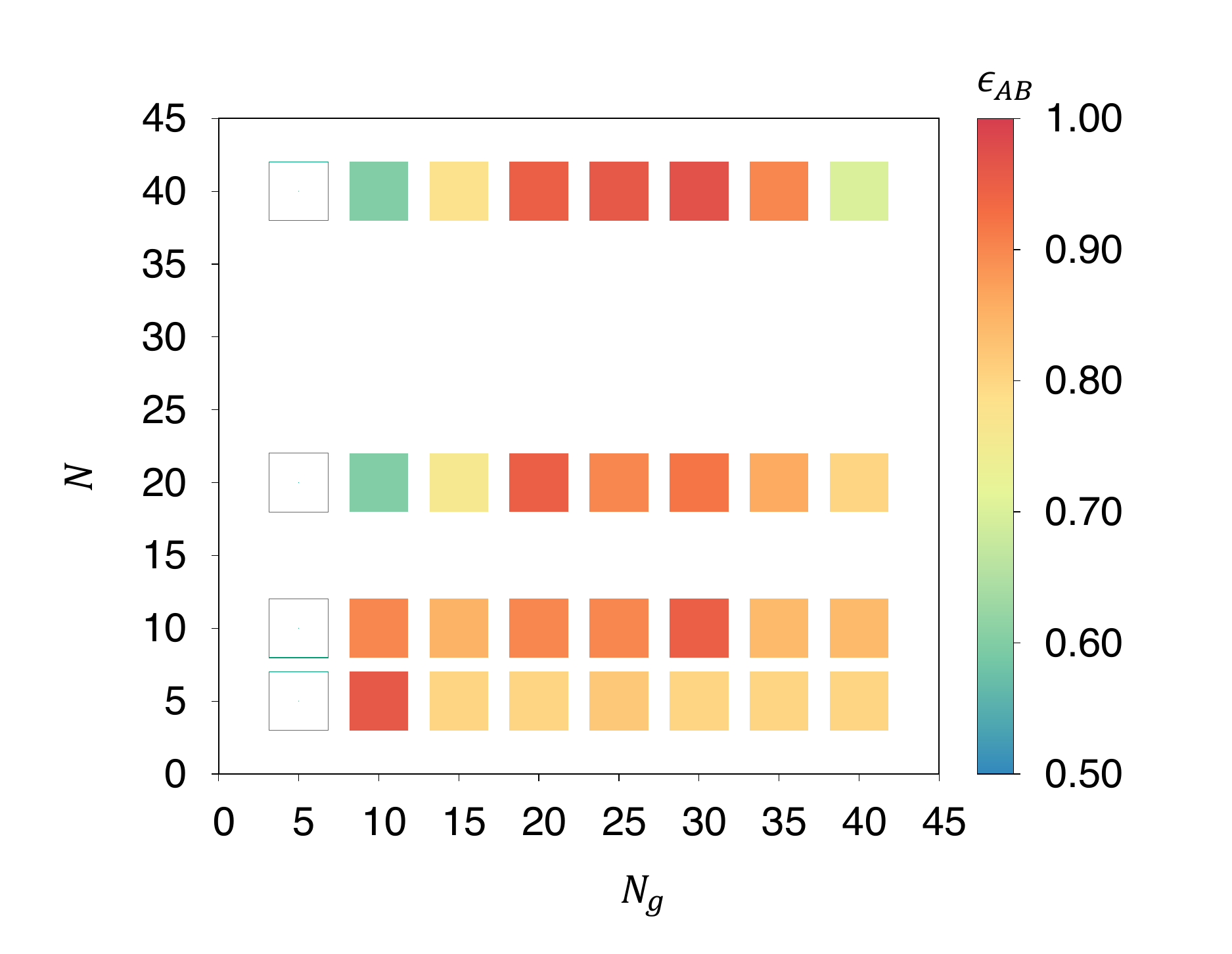}
\caption{Phase diagrams of PGNPs for the case $D=1\sigma$ (star polymers, upper panel) and $D=4\sigma$ 
(lower panel) as a function of the chain length, $N$, and the number of grafted chains, $N_{\rm g}$. 
Open squares indicate a disordered phase for all $\epsilon_{AB}$ values explored in 
this work. The colour bar denotes the ODT by means
of the potential interaction $\epsilon_{AB}$.}
\label{Fig2}
\end{figure}

\section{Results and Discussion}
%\vspace{0.5cm}
%\noindent
Our results obtained by MD simulations are 
summarised in Fig.~\ref{Fig2}, where the phase 
diagrams for two different cases are illustrated ($D=1\sigma$, 
upper panel
and $4\sigma$, lower panel ) as a function of the polymer chain length
$N$ and the number of grafted chains $N_{\rm g}$. The
colour bar indicates the value of the parameter $\epsilon_{\rm AB}$ for which
PGNPs with different type of grafted polymer chains can separate adopting
morphologies such as the ones shown in Fig.~\ref{Fig1}.
Lower values of $\epsilon_{\rm AB}$ indicate higher demixing difficulty, whereas
larger values of $\epsilon_{\rm AB}$ indicate much easier demixing of the 
PGNPs. The $\epsilon_{\rm AB}$ values
of Fig.~\ref{Fig2}
are reported as Supplementary Information. Figure~\ref{Fig1} also illustrates examples of mixed (disordered, middle panel) and
demixed (ordered, right panel) states. Therefore, the value of $\epsilon_{\rm AB}$ relates
to the ODT of our systems.
The phase diagram for $D=7\sigma$ is not presented here, 
as no phase separation between PGNPs with chains
of type A and B could be achieved within the available simulation time
for the range of $N$ and $N_{\rm g}$ values considered, 
in agreement with previous studies\cite{Chremos2011}.
In the case of $D=7\sigma$, we were able to detect only disordered configurations such as the
ones illustrated in Fig.~\ref{Fig1} (middle bottom panel).
Overall, we have found that the increase of the core particle hinders the 
phase separation of the systems for given $N$ and $N_{\rm g}$. Moreover, our results suggest
that there is a threshold for the size of the core particle that prevents
demixing of the PGNPs.  For a given
size of the core particle, the dependence on $N_{\rm g}$ and $N$ is not trivial (\textit{e.g.} monotonic),
and will be discussed below for two different cases of NP diameter, $D$, namely for the cases $D=1\sigma$ and $4\sigma$.

The case $D=1\sigma$ (Fig.~\ref{Fig2}, upper panel)
indicates that the phase separation takes place easier when the
length of the polymer chains is small (\textit{e.g.}, $N=5$,
Fig.~\ref{Fig2}, upper panel). In this case, values of
$\epsilon_{\rm AB} \approx 0.98$ can induce a phase separation between PGNPs with
polymer chains of different type. Hence, a very small incompatibility
between the different chemical beads of the grafted polymer chains leads to
PGNPs segregation. Moreover, differences
in the grafted density seem to play a minor role in 
the ODT for small chain length, $N$, which indicates that PGNPs behave like
small LJ beads with a similar effective diameter for the range of $N_{\rm g}$ considered here (Fig.~\ref{Fig2}, upper panel). 
However, as the chain length, $N$, increases the dependence on $N_{g}$ is more pronounced
and the resulting behaviour varies, depending on the value of $N$. 
For example, in the case of $N=20$ an increase of the grafting density $N_{\rm g}$
leads to a much easier demixing, a trend which is opposite for $N=40$
(Fig.~\ref{Fig2}, upper panel), which 
is an effect of the interplay between $N_{g}$ and $N$.
Any further conclusions at this point in the case of $D=1\sigma$ would
require a much larger range of $N_g$, which is not possible
for $D=1\sigma$. Therefore, we continue our discussion with the 
$D=4\sigma$ case.

In the case of $D=4\sigma$ (Fig.~\ref{Fig2}, lower panel), 
we observe that a smaller number of grafted chains
($N_{g}$), which exposes the steric interaction between the core particles (when $D=7\sigma$, even $N_g$ as large as 40 is not able to
cover the NP as is illustrated in the left panel of Fig.~\ref{Fig1})
and minimises the interaction between the grafted polymer chains of different type, 
hinders the demixing of PGNPs. For example, PGNPs with $N_{\rm g}=5$ and 
chain lengths as long as $N=40$ cannot shield the steric interaction between
the NP cores (Fig.~\ref{Fig2}, lower panel). Hence, for $N_{\rm g}=5$ no phase separation occurs 
for any of the $N$ values considered in this study.
The most interesting behaviour is observed when
we keep $N$ constant and vary $N_{\rm g}$ as the tethered polymer chains start to
occupy the NP's surface with attractive beads. If we consider the case $N=5$, then $N_{\rm g}=10$ 
indicates easy segregation of the PGNPs (Fig.~\ref{Fig2}, lower panel). However, further increase of $N_{\rm g}$ leads to a
higher stability of the melt, which seems to be
independent of the number of the attached polymer chains. In the latter case, the increase in $N_{\rm g}$ 
leads to an increase of the effective steric core interaction of the PGNPs \cite{Chremos2015},
which
eventually results in more difficult demixing, due to
the increase of the effective core size of the PGNPs. In the case of $N=10$, the change in
behaviour with increasing $N_{\rm g}$ takes place within a broader range of $N_{\rm g}$ values, 
due to the
higher value of the chain length $N$. As we increase the chain length $N$, the dependence on
increasing values of $N_{\rm g}$ is similar (Fig.~\ref{Fig2}, lower panel). 
A detailed analysis of our data yields the 
following key points:
On the one hand, the increase of the number of grafted chains, $N_g$, results 
initially in better demixing between PGNPs of different type (A- or B-type of
polymer chains). On the other hand, further increase of $N_g$ leads
to an effective increase in the size of NPs, which in turn hinders the phase
separation of the PGNPs. The interplay between these two effects depends
on the size of the NPs, $D$, and the chain length, $N$, of the grafted polymers.
In general,  the increase of the chain length favours phase separation, as
expected ($\chi_{\rm eff} ~\sim N_{\rm eff}$). These parameters determine
the value of $\epsilon_{\rm AB}$ where segregation 
of A- and B-type PGNPs  will take place.

\begin{figure}[t]
\includegraphics[width=85mm]{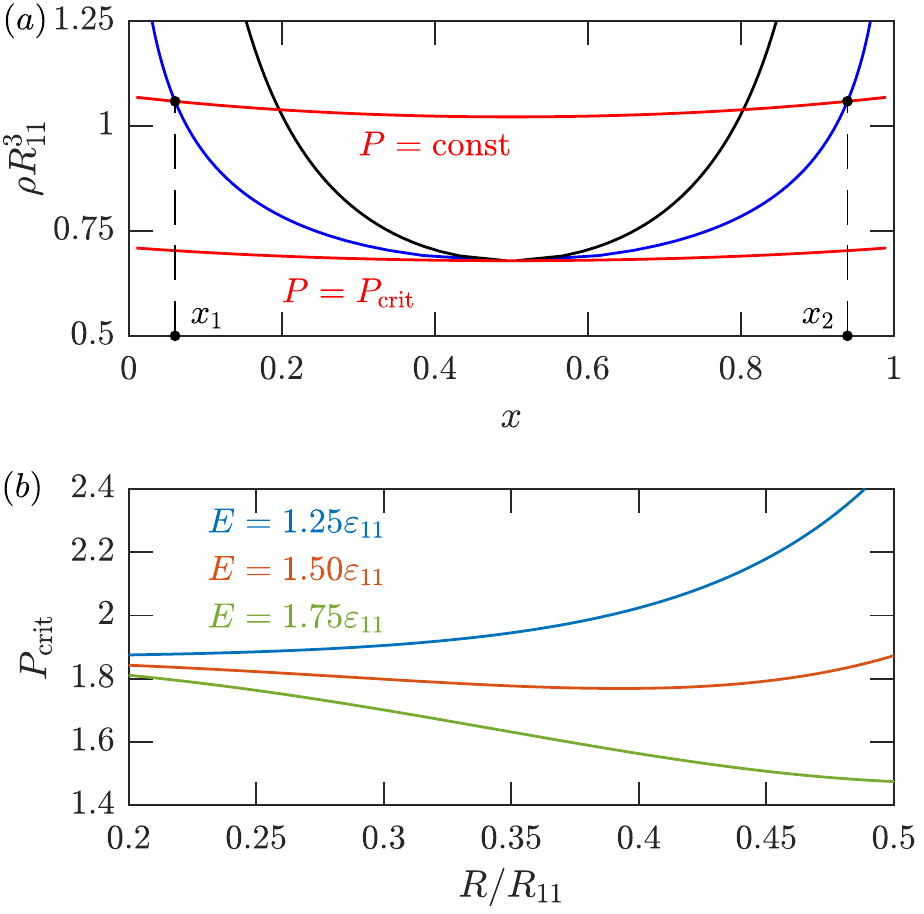}
\caption{The effect of growing core on the critical pressure of mixing--demixing
transition. (a) Binodal (blue) and spinodal (black) of mixing--demixing transition in the model
fluid. The coexisting phases with molar fractions $x_1$ and $x_2$ lie at the intersection of 
the binodal with the isochore $P = const$ (red). The demixing transition only happens above
$P=P_{\rm crit}$. (b) The dependence of $P_{\rm crit}$ on the radius of the hard-sphere repulsion for
different coupling between the hard-sphere and soft repulsion.
\label{Fig_P1}}
\end{figure}

We can try to explain the above effect with the help of a simple mean-field model, which coarse-grains the complicated interactions between PGNPs in the melt. As prompted, \textit{e.g.} by the first column of Fig.~\ref{Fig1}, the increase of the polymer chain lengths leads to the effective coating of the PGNP actual core particle in the melt, to the point where the entire melt can be approximated as being made up from effective `particles' with larger cores of different sizes and rigidity. These effective `particles' represent our level of coarse-graining. Still, the interactions between PGNPs are too complex, because the cores depend non-trivially on the grafting densities and chain lengths, and we would like to introduce further simplifications by splitting the interaction into simple parts:
\begin{equation}
    \varphi_{ij}^{total}(r)=\varphi_{ij}^{pol}(r)+\varphi^{hs}(r)+\varphi_{ij}^{soft}(r),
\end{equation}
where $i$ and $j$ index the polymer species. Since each PGNP is formed by grafting a hard particle with softly repulsive polymer chains, our effective interaction must include the potential of polymer chains. Such soft inter-chain repulsions are usually modelled by the generalized Gaussian-core potential: \cite{LouisBolhuisEtAlPhysRevE2000, ArcherEvansPhysRevE2001}
\begin{equation}
    \varphi_{ij}^{pol}(r)=\varepsilon_{ij}\exp{(-r^2/R^2_{ij})}.
\label{a1}    
\end{equation}
In polymer mixtures, $R_{ij}$ correspond to the gyration radii of the components (with $R_{12}=\sqrt{(R_{11}^2+R_{22}^2)/2}$), and $\varepsilon_{ij}$ models the repulsion strength. In our case, the chains are attached to the PGNP core particle, but the chain-induced interactions must be present in the model. Therefore, we still associate $R_{ij}$ with the chain length ($N$), but the strength of interaction should obviously be related to the number of chains, attached to the same PGNP core ($N_g$). The next component of the coarse-grained interactions is hard repulsion, caused by the tight wrapping of the polymer chains around the PGNP core. We will model it as a hard sphere of some radius $R$ (where $R\geq D/2$), noting that it is reasonable to expect that higher grafting densities should lead to larger values of $R$:
\begin{equation}
    \varphi^{hs}(r)=
    \begin{dcases}
    0, & \text{if } r>R\\
    \infty, & \text{if } r\leq R.
    \end{dcases}
\label{hs}
\end{equation}
Lastly, we need to account for the soft interactions, just outside of the hard core. These are caused by the "loose" parts of the wrapped polymer chains. For simplicity, we use the same form of the soft potential as we did for the free chains:
\begin{equation}
    \varphi_{ij}^{soft}(r)=E_{ij}\exp{(-r^2/R^2)}.
\label{a2}    
\end{equation}
The resulting bulk free energy per particle of the mixture as a function 
of the molar fraction $x$ and total density $\rho$ is given by the following expression
\begin{equation}
f(x,\rho)=f_{\rm id}(x,\rho)+f_{\rm hs}(x,\rho)+f_{\rm mf}(x,\rho),
\end{equation}
where the ideal part $f_{\rm id}(x,\rho)$ is defined as
\begin{equation}
f_{\rm id}(x,\rho)=k_B T (x \log{x}+(1-x) \log{(1-x)}+ \log{\rho}-1),
\end{equation}
the hard-sphere part is modelled using the Percus--Yevick equation of state, 
which in our case is independent of $x$, due to symmetry
\begin{equation}
f_{\rm hs}(x,\rho)=k_B T \left(\frac{3 (2-\eta) \eta}{2 \left(1-\eta\right)^2}-\log{(1-\eta)}\right), \quad \eta=\frac{4}{3}\pi R^3 \rho,
\end{equation}
and the soft repulsive part due to $\varphi_{ij}^{pol}(r)+\varphi_{ij}^{soft}(r)$ is given by the mean-field term \cite{ArcherEvansPhysRevE2001}
\begin{equation}
f_{\rm mf}(x,\rho)=\frac{1}{2}\rho\left((1-x)^2 V_{11}+2 x (1-x) V_{12}+x^2 V_{22}\right),
\end{equation}
with $V_{ij}$ being the total integrated strength of the potential: 
\begin{equation}
V_{ij}=\pi^{3/2} \left(R_{ij}^3\varepsilon_{ij}+R^3 E_{ij}\right),
\end{equation}
where due to symmetry we can set $E_{11}=E_{22}=1$, leaving us with $E_{12}=E$ as the coupling parameter. The mean-field treatment we follow supposes that the hard-sphere particles are moving in a uniform repulsive field induced by the soft potential $\varphi_{ij}^{pol}(r)+\varphi_{ij}^{soft}(r)$. This approach is essentially the same as the one which underpins the celebrated van der Waals equation of state of simple liquids, \cite{HansenMcDonald2006} with a change that the perturbation in our case is given by soft repulsions. \cite{LouisBolhuisEtAlPhysRevE2000, ArcherEvansPhysRevE2001} We further chose a system of units where $R_{11}=R_{22}=1$ and $\varepsilon_{11}=\varepsilon_{22}=1$. To ensure phase separation at small values 
of $R$, we fix $\varepsilon_{12}=1.2\varepsilon_{11}$, and in order to enhance the 
effects even further we set $T=0.4\varepsilon_{11}$. 

The picture of phase coexistence is summarised in Fig.~\ref{Fig_P1}(a). 
Here the densities and molar fractions of the coexisting demixed phases are found as the
intersections of the binodal curve (blue) with the isochores $\rho(x)$: $P=$const (red). 
The mixing/demixing transition is accessible above the critical pressure $P_{\rm crit}$. 
In addition, we show the boundaries of stability of the demixed phases -- the spinodal 
(black curve). A qualitative agreement with the results in Fig.~\ref{Fig2}
is expressed by the dependence of the critical pressure on the hard core radius $R$, 
plotted in Fig.~\ref{Fig_P1}(b) for several values of the coupling parameter. 
There are three possible scenarios which $P_{\rm crit}(R)$ may follow. 
For weak and strong coupling ($E=1.25$ and $E=1.75$), the dependence of $P_{\rm crit}(R)$, 
is monotonic. This means that increasing $N_{\rm g}$ should make the transition respectively 
less and more accessible. Notice the existence of the intermediate regime at $E=1.50$, 
where $P_{\rm crit}(R)$ has a local minimum. In this case, increasing $N_{\rm g}$ should initially facilitate
phase separation (by reducing $P_{\rm crit}(R)$), but further increase of $N_{\rm g}$ should actually inhibit
phase separation, as  $P_{\rm crit}$ starts to grow.

\begin{figure}[t]
\vspace{2mm}
\includegraphics[width=90mm]{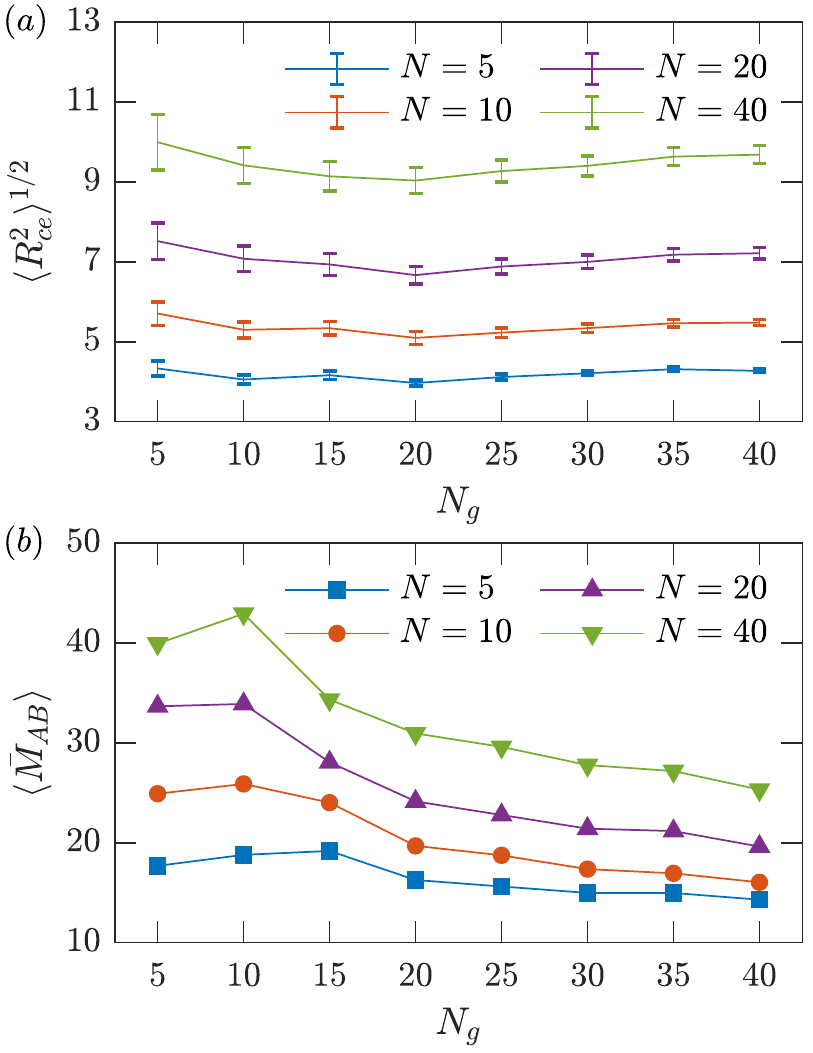}
\caption{For the case $D=4\sigma$ and different chain lengths $N$, we plot as functions of $N_{\rm g}$: (a) End-to-end distance of grafted chains (with error bars), and (b) Average number of neighbours for each PGNP.}
\label{Fig4}
\end{figure}

%\begin{figure}[t]
%\vspace{2mm}
%\includegraphics[width=90mm]{Fig4.pdf}
%\caption{End-to-end distance of grafted %chains vs. $N_{\rm g}$ for 
%the case $D=4\sigma$ and different chain %lengths $N$, as indicated.}
%\label{Fig3}
%\end{figure}

%\begin{figure}[t]
%\vspace{2mm}
%\includegraphics[width=90mm]{Fig5.pdf}
%\caption{Average number of neighbours for %each PGNP 
%vs. $N_{\rm g}$ for the case $D=4\sigma$
%and different chain lengths $N$, as %indicated.}
%\label{Fig4}
%\end{figure}

We have also analysed various structural properties from MD simulations
that contain information on the
interplay between the growing with $N_{\rm g}$ effective core size and the
growing with $N$ effective chain length of
the polymers. 
All our results are for the case $\epsilon_{\rm AB}=1.0$ in order to 
isolate the structural components that may play a role in the phase separation of PGNPs.
Figure~\ref{Fig4}(a) presents results for the end-to-end distance of the polymer chains,
\textit{i.e.}, the distance between the bead attached onto the core NP 
and the free end of each polymer chain, ${\langle R_{ce}^2 \rangle}^{1/2}$. An average over
all chains for each PGNP and an ensemble average is taken. Results
for the most interesting case $D=4\sigma$ are presented in Fig.~\ref{Fig4}(a), where more data is also available as a wider range 
of $N_{\rm g}$ is possible. 
Our results suggest a trend that the chain size is larger in the
case of systems that undergo easier demixing (e.g. $N_{\rm g}>15$), 
which may  suggest that grafted polymer chains
be closer to the NPs and the core effect be more pronounced.
Moreover, an increase in the average number of neighbours (Fig.~\ref{Fig4}(b)),
$\langle \overline M_{\rm AB} \rangle$,
is observed for small values of $N_{\rm g}$, which may indicate a larger role of
the effective core interactions 
in this case, in agreement with the trend observed in the case
of ${\langle R_{ce}^2 \rangle}^{1/2}$. 
In particular, $\langle \overline M_{\rm AB} \rangle$ overall decreases for each
$N$ when $N_{\rm g}$ increases, but a peak appears  for small $N{\rm g}$.
This peak disappears
when the number of grafted chains $N_{\rm g}>15$. Here, PGNPs are
considered as neighbours when at least one of the beads belonging
to different PGNPs interact without attempting to
analyse possible differences in the morphologies of
polymer chains that may depend on the relative position/orientation of
neighbouring PGNPs \cite{Travesset2017,Waltmann2017}. The analysis of the average
number of neighbours based only on the colloids (ignoring the polymer chains)
did not show any sensitivity on the
variation with $N_{\rm g}$. 
Overall, ${\langle R_{ce}^2 \rangle}^{1/2}$ and $\langle \overline M_{\rm AB} \rangle$
show some correlation with the phase behaviour of Fig.~\ref{Fig2}, but a direct correlation
of both properties with the phase diagram should not be overstated. In general, we observe 
a larger variation of ${\langle R_{ce}^2 \rangle}^{1/2}$ with $N_{\rm g}$ for larger
values of $N$ and smaller chain dimensions for smaller $N_{\rm g}$ values. The smaller
${\langle R_{ce}^2 \rangle}^{1/2}$ combined with the increase 
in $\langle \overline M_{\rm AB} \rangle$
for $N_{\rm g}<20$  may suggest an increased 
influence of the effective core size of the PGNPs.

\section{Concluding Remarks}
%\vspace{0.5cm}
%\noindent
%\textit{Concluding Remarks}.
We have studied the mixing--demixing behaviour of molecularly and compositionally symmetric
melts with spherical NPs with different chemical types of
grafted chains by using MD simulations of a coarse-grained model. The size of the NP, 
the number of grafted chains, and the length of the side chains were varied in order to investigate
the phase behaviour of these systems. Our results indicate that the increase of the
NP size hinders the phase separation. Moreover, the system cannot separate beyond a 
certain NP size for given grafting density and length of polymer chains. 
The phase behaviour of intermediate NP sizes (\textit{i.e.}, $D=4\sigma$) indicates a
transition region as the grafting density increases and the steric repulsions are
occupied by an attractive core with larger effective diameter, which is formed
by the tethered polymer chains. 
We have explained these effects and provided an analytical description by means of a mean-field theory
model. We anticipate that our results will provide further insight into the phase behaviour of PGNPs,
guided by the interplay phenomena between the NP size, the chain length, and the grafting density as
illustrated in this work, with implications in materials science and medicine.

\section*{Conflicts of interest}
There are no conflicts to declare.

\section*{Acknowledgements}
The authors acknowledge useful discussions with Alexandros Chremos. 
This research was supported in part by PLGrid Infrastructure.

%%%END OF MAIN TEXT%%%

%The \balance command can be used to balance the columns on the final page if desired. It should be placed anywhere within the first column of the last page.

\balance

%If notes are included in your references you can change the title from 'References' to 'Notes and references' using the following command:
%\renewcommand\refname{Notes and references}

%%%REFERENCES%%%
\bibliography{rsc} %You need to replace "rsc" on this line with the name of your .bib file

\providecommand*{\mcitethebibliography}{\thebibliography}
\csname @ifundefined\endcsname{endmcitethebibliography}
{\let\endmcitethebibliography\endthebibliography}{}
\begin{mcitethebibliography}{39}
\providecommand*{\natexlab}[1]{#1}
\providecommand*{\mciteSetBstSublistMode}[1]{}
\providecommand*{\mciteSetBstMaxWidthForm}[2]{}
\providecommand*{\mciteBstWouldAddEndPuncttrue}
  {\def\EndOfBibitem{\unskip.}}
\providecommand*{\mciteBstWouldAddEndPunctfalse}
  {\let\EndOfBibitem\relax}
\providecommand*{\mciteSetBstMidEndSepPunct}[3]{}
\providecommand*{\mciteSetBstSublistLabelBeginEnd}[3]{}
\providecommand*{\EndOfBibitem}{}
\mciteSetBstSublistMode{f}
\mciteSetBstMaxWidthForm{subitem}
{(\emph{\alph{mcitesubitemcount}})}
\mciteSetBstSublistLabelBeginEnd{\mcitemaxwidthsubitemform\space}
{\relax}{\relax}

\bibitem[Balazs \emph{et~al.}(2006)Balazs, Emrick, and Russell]{Balazs2006}
A.~Balazs, T.~Emrick and T.~Russell, \emph{Science}, 2006, \textbf{314},
  1107--1110\relax
\mciteBstWouldAddEndPuncttrue
\mciteSetBstMidEndSepPunct{\mcitedefaultmidpunct}
{\mcitedefaultendpunct}{\mcitedefaultseppunct}\relax
\EndOfBibitem
\bibitem[Choudhury \emph{et~al.}(2015)Choudhury, Agrawal, Kim, and
  Archer]{Choudhury2015}
S.~Choudhury, A.~Agrawal, S.~A. Kim and L.~A. Archer, \emph{Langmuir}, 2015,
  \textbf{31}, 3222--3231\relax
\mciteBstWouldAddEndPuncttrue
\mciteSetBstMidEndSepPunct{\mcitedefaultmidpunct}
{\mcitedefaultendpunct}{\mcitedefaultseppunct}\relax
\EndOfBibitem
\bibitem[Schaefer \emph{et~al.}(2011)Schaefer, Moganty, Yanga, and
  Archer]{Schaefer2011}
J.~L. Schaefer, S.~S. Moganty, D.~A. Yanga and L.~A. Archer, \emph{J. Mater.
  Chem.}, 2011, \textbf{21}, 10094--10102\relax
\mciteBstWouldAddEndPuncttrue
\mciteSetBstMidEndSepPunct{\mcitedefaultmidpunct}
{\mcitedefaultendpunct}{\mcitedefaultseppunct}\relax
\EndOfBibitem
\bibitem[Hooper and Schweizer(2005)]{Hooper2005}
J.~B. Hooper and K.~S. Schweizer, \emph{Macromolecules}, 2005, \textbf{38},
  8858--8869\relax
\mciteBstWouldAddEndPuncttrue
\mciteSetBstMidEndSepPunct{\mcitedefaultmidpunct}
{\mcitedefaultendpunct}{\mcitedefaultseppunct}\relax
\EndOfBibitem
\bibitem[Hooper and Schweizer(2006)]{Hooper2006}
J.~B. Hooper and K.~S. Schweizer, \emph{Macromolecules}, 2006, \textbf{39},
  5133--5142\relax
\mciteBstWouldAddEndPuncttrue
\mciteSetBstMidEndSepPunct{\mcitedefaultmidpunct}
{\mcitedefaultendpunct}{\mcitedefaultseppunct}\relax
\EndOfBibitem
\bibitem[Sunday \emph{et~al.}(2012)Sunday, Ilavsky, and Green]{Sunday2012}
D.~Sunday, J.~Ilavsky and D.~L. Green, \emph{Macromolecules}, 2012,
  \textbf{45}, 4007--4011\relax
\mciteBstWouldAddEndPuncttrue
\mciteSetBstMidEndSepPunct{\mcitedefaultmidpunct}
{\mcitedefaultendpunct}{\mcitedefaultseppunct}\relax
\EndOfBibitem
\bibitem[Srivastava \emph{et~al.}(2017)Srivastava, Choudhury, Agrawal, and
  Archer]{Srivastava2017}
S.~Srivastava, S.~Choudhury, A.~Agrawal and L.~A. Archer, \emph{Curr. Opin.
  Chem. Eng.}, 2017, \textbf{16}, 92--101\relax
\mciteBstWouldAddEndPuncttrue
\mciteSetBstMidEndSepPunct{\mcitedefaultmidpunct}
{\mcitedefaultendpunct}{\mcitedefaultseppunct}\relax
\EndOfBibitem
\bibitem[Srivastava \emph{et~al.}(2014)Srivastava, Schaefer, Yang, Tu, and
  Archer]{Srivastava2014}
S.~Srivastava, J.~Schaefer, Z.~Yang, Z.~Tu and L.~Archer, \emph{Adv. Mater.},
  2014, \textbf{15}, 201--234\relax
\mciteBstWouldAddEndPuncttrue
\mciteSetBstMidEndSepPunct{\mcitedefaultmidpunct}
{\mcitedefaultendpunct}{\mcitedefaultseppunct}\relax
\EndOfBibitem
\bibitem[Mangal \emph{et~al.}(2015)Mangal, Srivastava, and Archer]{Mangal2015}
R.~Mangal, S.~Srivastava and L.~Archer, \emph{Nat. Commun.}, 2015, \textbf{6},
  7198\relax
\mciteBstWouldAddEndPuncttrue
\mciteSetBstMidEndSepPunct{\mcitedefaultmidpunct}
{\mcitedefaultendpunct}{\mcitedefaultseppunct}\relax
\EndOfBibitem
\bibitem[Srivastava \emph{et~al.}(2012)Srivastava, Agarwal, and
  Archer]{Srivastava2012}
S.~Srivastava, P.~Agarwal and L.~Archer, \emph{Langmuir}, 2012, \textbf{28},
  6272--6281\relax
\mciteBstWouldAddEndPuncttrue
\mciteSetBstMidEndSepPunct{\mcitedefaultmidpunct}
{\mcitedefaultendpunct}{\mcitedefaultseppunct}\relax
\EndOfBibitem
\bibitem[Agrawal \emph{et~al.}(2016)Agrawal, Wenning, Choudhury, and
  Archer]{Agrawal2016}
A.~Agrawal, B.~M. Wenning, S.~Choudhury and L.~A. Archer, \emph{Langmuir},
  2016, \textbf{32}, 8698--8708\relax
\mciteBstWouldAddEndPuncttrue
\mciteSetBstMidEndSepPunct{\mcitedefaultmidpunct}
{\mcitedefaultendpunct}{\mcitedefaultseppunct}\relax
\EndOfBibitem
\bibitem[Chremos and Panagiotopoulos(2011)]{Chremos2011}
A.~Chremos and A.~Z. Panagiotopoulos, \emph{Phys. Rev. Lett.}, 2011,
  \textbf{107}, 105503\relax
\mciteBstWouldAddEndPuncttrue
\mciteSetBstMidEndSepPunct{\mcitedefaultmidpunct}
{\mcitedefaultendpunct}{\mcitedefaultseppunct}\relax
\EndOfBibitem
\bibitem[Chremos \emph{et~al.}(2011)Chremos, Panagiotopoulos, Yu, and
  Koch]{Chremos2011b}
A.~Chremos, A.~Z. Panagiotopoulos, H.~Y. Yu and D.~L. Koch, \emph{J. Chem.
  Phys.}, 2011, \textbf{135}, 114901\relax
\mciteBstWouldAddEndPuncttrue
\mciteSetBstMidEndSepPunct{\mcitedefaultmidpunct}
{\mcitedefaultendpunct}{\mcitedefaultseppunct}\relax
\EndOfBibitem
\bibitem[Chremos and Douglas(2016)]{Chremos2016}
A.~Chremos and J.~F. Douglas, \emph{Soft Matter}, 2016, \textbf{12},
  9527--9537\relax
\mciteBstWouldAddEndPuncttrue
\mciteSetBstMidEndSepPunct{\mcitedefaultmidpunct}
{\mcitedefaultendpunct}{\mcitedefaultseppunct}\relax
\EndOfBibitem
\bibitem[Chremos and Douglas(2017)]{Chremos2017}
A.~Chremos and J.~F. Douglas, \emph{Ann. Phys.}, 2017, \textbf{529},
  1600342\relax
\mciteBstWouldAddEndPuncttrue
\mciteSetBstMidEndSepPunct{\mcitedefaultmidpunct}
{\mcitedefaultendpunct}{\mcitedefaultseppunct}\relax
\EndOfBibitem
\bibitem[Yu and Koch(2010)]{Yu2010}
H.-Y. Yu and D.~L. Koch, \emph{Langmuir}, 2010, \textbf{26}, 16801--16811\relax
\mciteBstWouldAddEndPuncttrue
\mciteSetBstMidEndSepPunct{\mcitedefaultmidpunct}
{\mcitedefaultendpunct}{\mcitedefaultseppunct}\relax
\EndOfBibitem
\bibitem[Yu \emph{et~al.}(2014)Yu, Srivastava, Archer, and Koch]{Yu2014}
H.-Y. Yu, S.~Srivastava, L.~A. Archer and D.~L. Koch, \emph{Soft Matter}, 2014,
  \textbf{10}, 9120--9135\relax
\mciteBstWouldAddEndPuncttrue
\mciteSetBstMidEndSepPunct{\mcitedefaultmidpunct}
{\mcitedefaultendpunct}{\mcitedefaultseppunct}\relax
\EndOfBibitem
\bibitem[Agarwal \emph{et~al.}(2010)Agarwal, Qi, and Archer]{Agarwal2010}
P.~Agarwal, H.~Qi and L.~A. Archer, \emph{Nano Lett.}, 2010, \textbf{10},
  111--115\relax
\mciteBstWouldAddEndPuncttrue
\mciteSetBstMidEndSepPunct{\mcitedefaultmidpunct}
{\mcitedefaultendpunct}{\mcitedefaultseppunct}\relax
\EndOfBibitem
\bibitem[Fernandes \emph{et~al.}(2013)Fernandes, Koerner, Giannelis, and
  Vaia]{Fernandes2013}
N.~J. Fernandes, H.~Koerner, E.~P. Giannelis and R.~A. Vaia, \emph{MRS
  Commun.}, 2013, \textbf{3}, 13--29\relax
\mciteBstWouldAddEndPuncttrue
\mciteSetBstMidEndSepPunct{\mcitedefaultmidpunct}
{\mcitedefaultendpunct}{\mcitedefaultseppunct}\relax
\EndOfBibitem
\bibitem[Agrawal \emph{et~al.}(2015)Agrawal, Yu, Srivastava, Choudhury,
  Narayanan, and Archer]{Agrawal2015}
A.~Agrawal, H.-Y. Yu, S.~Srivastava, S.~Choudhury, S.~Narayanan and L.~A.
  Archer, \emph{Soft Matter}, 2015, \textbf{11}, 5224--5234\relax
\mciteBstWouldAddEndPuncttrue
\mciteSetBstMidEndSepPunct{\mcitedefaultmidpunct}
{\mcitedefaultendpunct}{\mcitedefaultseppunct}\relax
\EndOfBibitem
\bibitem[Kim \emph{et~al.}(2015)Kim, Mangal, and Archer]{Kim2015}
S.~A. Kim, R.~Mangal and L.~A. Archer, \emph{Macromolecules}, 2015,
  \textbf{48}, 6280--6293\relax
\mciteBstWouldAddEndPuncttrue
\mciteSetBstMidEndSepPunct{\mcitedefaultmidpunct}
{\mcitedefaultendpunct}{\mcitedefaultseppunct}\relax
\EndOfBibitem
\bibitem[Agarwal \emph{et~al.}(2012)Agarwal, Kim, and Archer]{Agarwal2012}
P.~Agarwal, S.~A. Kim and L.~A. Archer, \emph{Phys. Rev. Lett.}, 2012,
  \textbf{109}, 258301\relax
\mciteBstWouldAddEndPuncttrue
\mciteSetBstMidEndSepPunct{\mcitedefaultmidpunct}
{\mcitedefaultendpunct}{\mcitedefaultseppunct}\relax
\EndOfBibitem
\bibitem[Kim and Archer(2014)]{Kim2014}
S.~A. Kim and L.~A. Archer, \emph{Macromolecules}, 2014, \textbf{47},
  687--694\relax
\mciteBstWouldAddEndPuncttrue
\mciteSetBstMidEndSepPunct{\mcitedefaultmidpunct}
{\mcitedefaultendpunct}{\mcitedefaultseppunct}\relax
\EndOfBibitem
\bibitem[Choudhury \emph{et~al.}(2015)Choudhury, Mangal, Agrawal, and
  Archer]{Choudhury2015b}
S.~Choudhury, R.~Mangal, A.~Agrawal and L.~A. Archer, \emph{Nat. Commun.},
  2015, \textbf{6}, 10101\relax
\mciteBstWouldAddEndPuncttrue
\mciteSetBstMidEndSepPunct{\mcitedefaultmidpunct}
{\mcitedefaultendpunct}{\mcitedefaultseppunct}\relax
\EndOfBibitem
\bibitem[Goyal and Escobedo(2011)]{Goyal2011}
S.~Goyal and F.~A. Escobedo, \emph{J. Chem. Phys.}, 2011, \textbf{135},
  184902\relax
\mciteBstWouldAddEndPuncttrue
\mciteSetBstMidEndSepPunct{\mcitedefaultmidpunct}
{\mcitedefaultendpunct}{\mcitedefaultseppunct}\relax
\EndOfBibitem
\bibitem[Agarwal \emph{et~al.}(2011)Agarwal, Srivastava, and
  Archer]{Agarwal2011}
P.~Agarwal, S.~Srivastava and L.~A. Archer, \emph{Phys. Rev. Lett.}, 2011,
  \textbf{107}, 268302\relax
\mciteBstWouldAddEndPuncttrue
\mciteSetBstMidEndSepPunct{\mcitedefaultmidpunct}
{\mcitedefaultendpunct}{\mcitedefaultseppunct}\relax
\EndOfBibitem
\bibitem[Meyer(1940)]{Meyer1940}
K.~Meyer, \emph{Helv. Chim. Acta}, 1940, \textbf{23}, 1063--1070\relax
\mciteBstWouldAddEndPuncttrue
\mciteSetBstMidEndSepPunct{\mcitedefaultmidpunct}
{\mcitedefaultendpunct}{\mcitedefaultseppunct}\relax
\EndOfBibitem
\bibitem[Flory(1953)]{Flory:1953/a}
P.~J. Flory, \emph{Principles of Polymer Chemistry}, Cornell University Press,
  Ithaca, NY, 1953\relax
\mciteBstWouldAddEndPuncttrue
\mciteSetBstMidEndSepPunct{\mcitedefaultmidpunct}
{\mcitedefaultendpunct}{\mcitedefaultseppunct}\relax
\EndOfBibitem
\bibitem[Chremos \emph{et~al.}(2014)Chremos, Nikoubashman, and
  Panagiotopoulos]{Chremos:2014/a}
A.~Chremos, A.~Nikoubashman and A.~Z. Panagiotopoulos, \emph{J. Chem. Phys.},
  2014, \textbf{140}, 054909\relax
\mciteBstWouldAddEndPuncttrue
\mciteSetBstMidEndSepPunct{\mcitedefaultmidpunct}
{\mcitedefaultendpunct}{\mcitedefaultseppunct}\relax
\EndOfBibitem
\bibitem[Kremer and Grest(1990)]{Kremer1990}
K.~Kremer and G.~S. Grest, \emph{J. Chem. Phys.}, 1990, \textbf{92}, 5057\relax
\mciteBstWouldAddEndPuncttrue
\mciteSetBstMidEndSepPunct{\mcitedefaultmidpunct}
{\mcitedefaultendpunct}{\mcitedefaultseppunct}\relax
\EndOfBibitem
\bibitem[Chremos and Theodorakis(2014)]{Chremos2014panos}
A.~Chremos and P.~E. Theodorakis, \emph{ACS Macro Lett.}, 2014, \textbf{3},
  1096--1100\relax
\mciteBstWouldAddEndPuncttrue
\mciteSetBstMidEndSepPunct{\mcitedefaultmidpunct}
{\mcitedefaultendpunct}{\mcitedefaultseppunct}\relax
\EndOfBibitem
\bibitem[Chremos and Theodorakis(2016)]{ChremosTheodorakis2016}
A.~Chremos and P.~E. Theodorakis, \emph{Polymer}, 2016, \textbf{97},
  191--195\relax
\mciteBstWouldAddEndPuncttrue
\mciteSetBstMidEndSepPunct{\mcitedefaultmidpunct}
{\mcitedefaultendpunct}{\mcitedefaultseppunct}\relax
\EndOfBibitem
\bibitem[Plimpton(1995)]{Plimpton:1995/a}
S.~J. Plimpton, \emph{J.~Comput.~Phys.}, 1995, \textbf{117}, 1--19\relax
\mciteBstWouldAddEndPuncttrue
\mciteSetBstMidEndSepPunct{\mcitedefaultmidpunct}
{\mcitedefaultendpunct}{\mcitedefaultseppunct}\relax
\EndOfBibitem
\bibitem[Chremos and Douglas(2015)]{Chremos2015}
A.~Chremos and J.~F. Douglas, \emph{J. Chem. Phys.}, 2015, \textbf{143},
  111104\relax
\mciteBstWouldAddEndPuncttrue
\mciteSetBstMidEndSepPunct{\mcitedefaultmidpunct}
{\mcitedefaultendpunct}{\mcitedefaultseppunct}\relax
\EndOfBibitem
\bibitem[Louis \emph{et~al.}(2000)Louis, Bolhuis, and
  Hansen]{LouisBolhuisEtAlPhysRevE2000}
A.~A. Louis, P.~Bolhuis and J.~P. Hansen, \emph{Phys. Rev. E}, 2000,
  \textbf{62}, 7961\relax
\mciteBstWouldAddEndPuncttrue
\mciteSetBstMidEndSepPunct{\mcitedefaultmidpunct}
{\mcitedefaultendpunct}{\mcitedefaultseppunct}\relax
\EndOfBibitem
\bibitem[Archer and Evans(2001)]{ArcherEvansPhysRevE2001}
A.~Archer and R.~Evans, \emph{Phys. Rev. E}, 2001, \textbf{64}, 041501\relax
\mciteBstWouldAddEndPuncttrue
\mciteSetBstMidEndSepPunct{\mcitedefaultmidpunct}
{\mcitedefaultendpunct}{\mcitedefaultseppunct}\relax
\EndOfBibitem
\bibitem[Hansen and McDonald(2006)]{HansenMcDonald2006}
J.~Hansen and I.~McDonald, \emph{Theory of {Simple} {Liquids} (3rd edn)},
  Elsevier, 2006\relax
\mciteBstWouldAddEndPuncttrue
\mciteSetBstMidEndSepPunct{\mcitedefaultmidpunct}
{\mcitedefaultendpunct}{\mcitedefaultseppunct}\relax
\EndOfBibitem
\bibitem[Travesset(2017)]{Travesset2017}
A.~Travesset, \emph{ACS Nano}, 2017, \textbf{11}, 5375--5382\relax
\mciteBstWouldAddEndPuncttrue
\mciteSetBstMidEndSepPunct{\mcitedefaultmidpunct}
{\mcitedefaultendpunct}{\mcitedefaultseppunct}\relax
\EndOfBibitem
\bibitem[Waltmann \emph{et~al.}(2017)Waltmann, Horst, and
  Travesset]{Waltmann2017}
C.~Waltmann, N.~Horst and A.~Travesset, \emph{ACS Nano}, 2017, \textbf{11},
  11273--11282\relax
\mciteBstWouldAddEndPuncttrue
\mciteSetBstMidEndSepPunct{\mcitedefaultmidpunct}
{\mcitedefaultendpunct}{\mcitedefaultseppunct}\relax
\EndOfBibitem
\end{mcitethebibliography}
\bibliographystyle{rsc} %the RSC's .bst file

\end{document}